\documentclass[a4paper,11pt]{article}

\usepackage{contribution}

\newcommand{\weblink}[2][]{%
    \ifthenelse{\equal{#1}{}}%
    {\textnormal{\url{#2}}}%
    {\textnormal{\href{#2}{#1}}}%
}

\def\beq{\begin{equation}}
\def\eeq#1{\label{#1}\end{equation}}
\def\eeqn{\end{equation}}

\def\beqa{\begin{eqnarray}}
\def\eeqa#1{\label{#1}\end{eqnarray}}
\def\eeqan{\end{eqnarray}}

\let\bar=\overbar

\def\Dslash{\not{\hbox{\kern-4pt $D$}}}
\def\dslash{\not{\hbox{\kern-2pt $\del$}}}

\def\msb{{\bar{\ssstyle M \kern -1pt S}}}

\newcommand{\contribution}[7][]{%
  \clearpage
  \thispagestyle{plain}
  \ifthenelse{\equal{#1}{}}
  {\hypersetup{pdftitle={#2}}}
  {\hypersetup{pdftitle={#1}}}
  \hypersetup{pdfauthor={{#3} {#4}}}
  {\centering\normalfont\LARGE\bfseries\sffamily #2 \par\nobreak}
  \lhead{}
  \chead{%
    \textit{\footnotesize XIV International Conference on Hadron Spectroscopy
      (\weblink[\textit{hadron2011}]{http://www.hadron2011.de}), 13-17 June 2011, Munich, Germany}%
  }
  \rhead{}
  \bigskip
  \begin{center}
    {#3} {#4}\ifthenelse{\equal{#6}{}}{}{\footnote{\weblink[#6]{mailto:#6}}}
    \ifthenelse{\equal{#7}{}}{}{#7} \\
    \textit{#5}
  \end{center}
  \bigskip
}

\renewcommand{\abstract}[1]{%
  \begin{center}
    \begin{minipage}{0.85\textwidth}
      \begin{footnotesize}
        #1
      \end{footnotesize}
    \end{minipage}
  \end{center}
  \bigskip
}

\begin{document}


%
%
%
%
%
{  


\newcommand{\PB}{\ensuremath{{\mathrm{B}}}}
\newcommand{\PaB}{\ensuremath{\mathrm{\overline{B}}}}
\newcommand{\bbar}{\ensuremath{{\overline{\mathrm{b}}}}\xspace}
\newcommand{\cbar}{\ensuremath{{\overline{\mathrm{c}}}}\, }


\newcommand{\bhadb}{{\PB\PaB\ }}
\newcommand{\bhadbbr}{{\PB\PaB\,}}

\newcommand{\Pbquark}{\ensuremath{\mathrm{b}}\xspace}
\newcommand{\Xbbbar}{\ensuremath{\mathrm{B\overline{B}}}\xspace}
\newcommand{\pt}{\ensuremath{p_{\mathrm{T}}}}
\newcommand{\ptrans}{\ensuremath{p_{\mathrm{T}}}}
\newcommand{\ptrel}{\ensuremath{p_{\mathrm{T,rel}}}}
\newcommand{\PYTHIA}{{\textsc{pythia}}}
\newcommand{\MADGRAPH} {\textsc{MadGraph}}
\newcommand{\CASCADE}{\textsc{cascade}}
\newcommand{\MCATNLO} {\textsc{mc@nlo}}
\newcommand{\ub} {\ensuremath{\mu {\mathrm{b}}}}
\newcommand{\ptmu} {\ensuremath{p_{T}^{\mu}}}
\newcommand{\etamu} {\ensuremath{\eta^{\mu}}}

%

\contribution[\PB\PaB\ Angular Correlations]  
{Measurement of \PB\PaB\ Angular Correlations \\
  at $\sqrt{s}=7$ TeV with the CMS Experiment}  
{Christoph}{Grab}  
{Institute for Particle Physics,   ETH Zurich,  \\
  CH-8093 Zurich, Switzerland}  
{}  
{on behalf of the CMS Collaboration}  
%

\abstract{%
Measurements of the angular correlations between beauty and 
anti-beauty hadrons produced in LHC pp collisions at $\sqrt{s}=7$ TeV are presented.
These results probe for the first time the small angular separation region and
show sensitivity to collinear particle emission.
The results are compared with predictions based on perturbative 
QCD calculations at leading and next-to-leading order.
}
%


\section{Introduction}

Studies of production properties of beauty quarks ($\mathrm{b}$) at the CERN LHC collider 
are of twofold interest.
Firstly, the $\mathrm{b}$ production process provides an excellent opportunity to study
details of perturbative Quantum Chromodynamics (pQCD). 
Over the years, the various tensions between the predictions and the measurements, 
that existed in data
at lower energies such as the HERA or the Tevatron collider, have been reduced, 
however not completely resolved. 
Studies at the LHC collider with higher centre-of-mass energies
complement the previous data, but also expand the reach and provide tests
at precisions below the present theoretical uncertainties.

Secondly, $\mathrm{b}$ quark production constitutes one of the major backgrounds
in many of the searches for new physics. Any production channel of 
exotic states, that produces top quarks or W-bosons, will inherently
have a large $\mathrm{b}$ production rate. 
It is of importance not only to understand the absolute production rates, but
also to be able to describe the details of the $\mathrm{b}$ production dynamics.
Thus, a solid understanding of the topology of the final states will be
crucial to constitute efficient criteria to distinguish possible signal 
signatures from $\mathrm{b}$-induced background configurations.

\begin{figure}[htb]
\centering
\begin{tabular}{cc}
\begin{minipage}{14cm}
\includegraphics[width=0.3\textwidth]{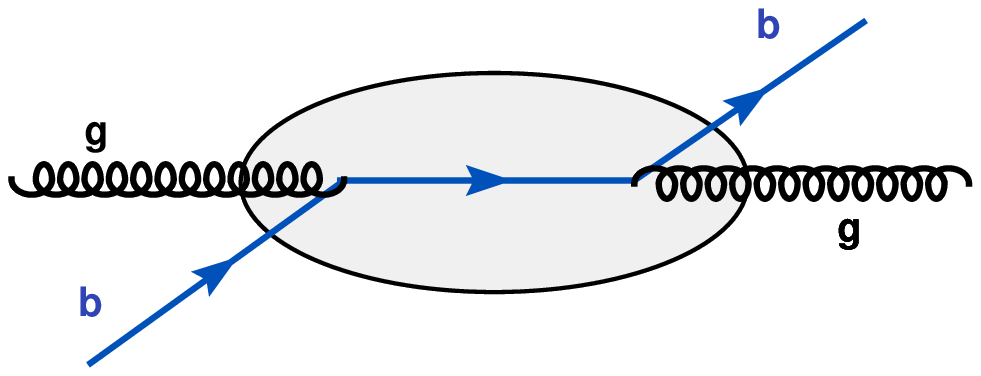} 
\includegraphics[width=0.3\textwidth]{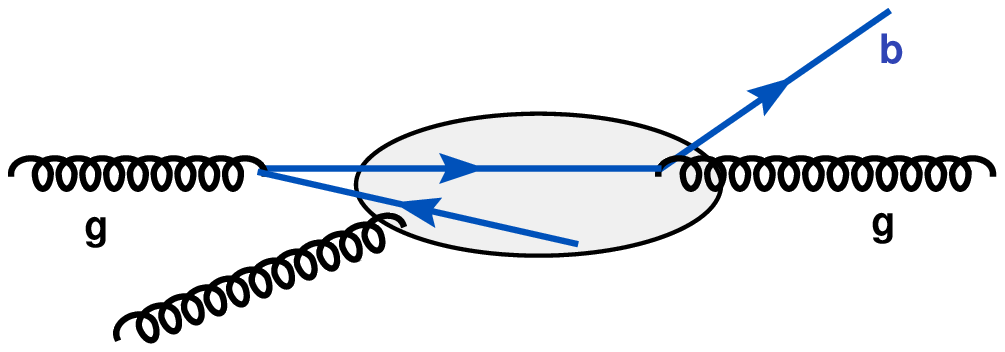} 
\includegraphics[width=0.3\textwidth]{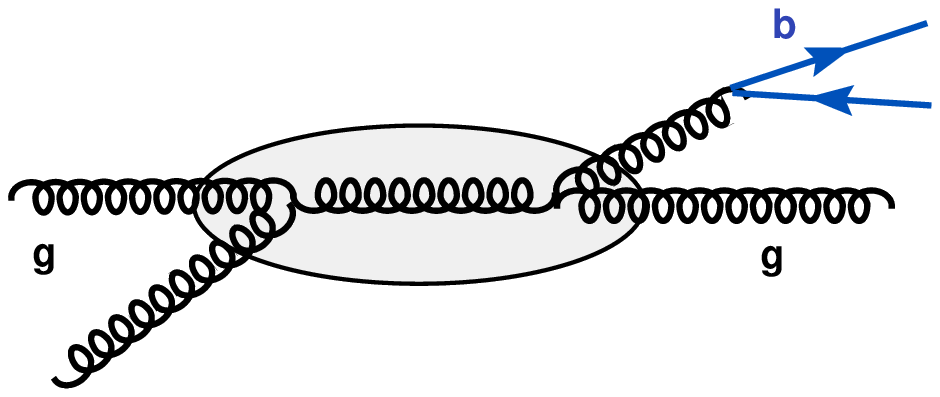} 
 \caption{ Examples of schematic Feynman diagrams for the three subprocesses: flavour creation 
(FCR, top), flavour excitation (FCR, middle) and gluon splitting (GSP, bottom).}
    \label{f:feyn}
\end{minipage}
\end{tabular}
\end{figure}

Within the leading-order (LO) QCD picture, the
production of $\mathrm{b\overline{b}}$  in $\mathrm{pp}$ collisions at LHC can
be attributed to three parton level production subprocesses, 
commonly denoted by flavour creation (FCR), 
flavour excitation (FEX) and gluon splitting (GSP) (see Fig.~\ref{f:feyn}). 
At higher orders, the distinction becomes
scale dependent, and  is thus less well defined.
Due to the different dynamics of these components, the final state topologies differ 
substantially from each other. 
FCR pertains to the $2\to 2$ processes gluon-fusion and $\mathrm{q\overline{q}}$ annihilation,
where the $\mathrm{b}$ and $\mathrm{\overline{b}}$ are emitted in a back-to-back configuration.
FEX refers to the $2\to 3$ process, where one $\mathrm{b}$ quark of a $\mathrm{b\overline{b}}$ pair 
from the proton sea participates in the hard scattering, thereby producing an asymmetry
in the momentum and angular distribution of the final state.
The GSP contribution on the other hand describes gluons from either initial or final state,
that split into a   $\mathrm{b\overline{b}}$ pair, which in turn are emitted
preferentially at small opening angles and low \pt.
Furthermore, the relative production rates themselves vary also as a function 
of the energy scale. It is expected, that at higher energies the 
gluon splitting contributions dominate, ie. processes with a collinear
branching of gluons into \bhadb pairs will become the major source of $\mathrm{b}$ quark
production.

\section{ Measurements of $\PB\PaB$ Angular Correlations }

CMS has performed the first measurement~\cite{Khachatryan:2011wq}
of the angular correlations between beauty and 
anti-beauty hadrons (\PB\PaB) produced in pp collisions at 
$ \sqrt{s} = 7\; \mbox{TeV}$,
thereby probing for the first time the region of small angular separation.
The analysis is based on a data sample corresponding 
to an integrated luminosity of $3.1\,\pm\, 0.3\, \mathrm{pb}^{-1}$.
A detailed description of the CMS detector can be found in 
Ref.~\cite{ref:2008zzk}.

The measurements are done differentially
as a function of the opening angle for different event scales,
which are characterised by the leading jet transverse momentum in the event
(independently of $\mathrm{b}$ hadrons).
The leading jet of the event is used to trigger. 
The trigger thresholds are chosen such as to reach an efficiency over 99\% for all three
energy scale bins, which correspond to a leading jet $\pt$ in excess
of 56, 84 and 120~GeV, respectively, when using corrected jet energies.

\begin{figure}[htb]
        \centering
\includegraphics[width=0.45\textwidth]{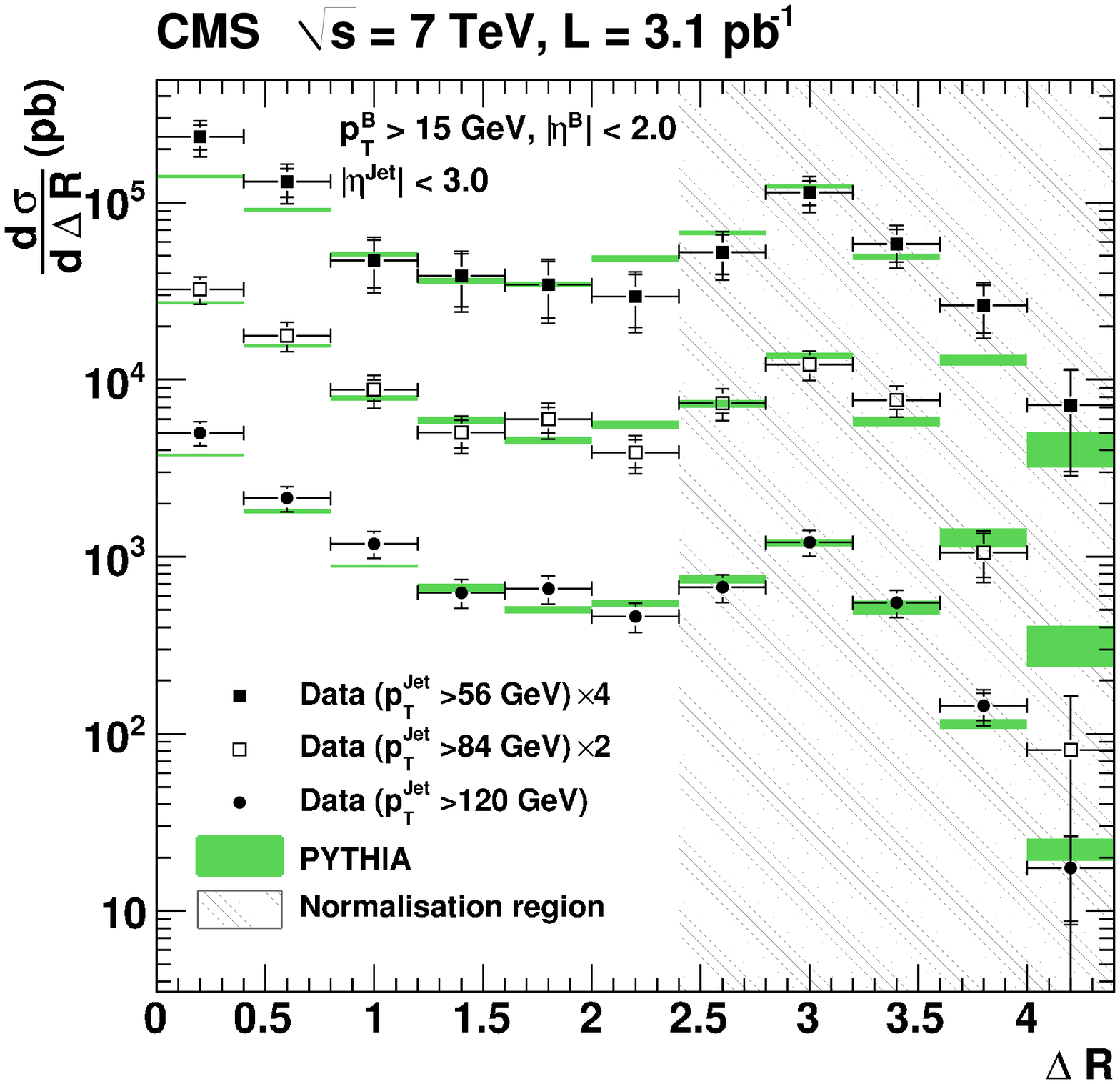}
\includegraphics[width=0.45\textwidth]{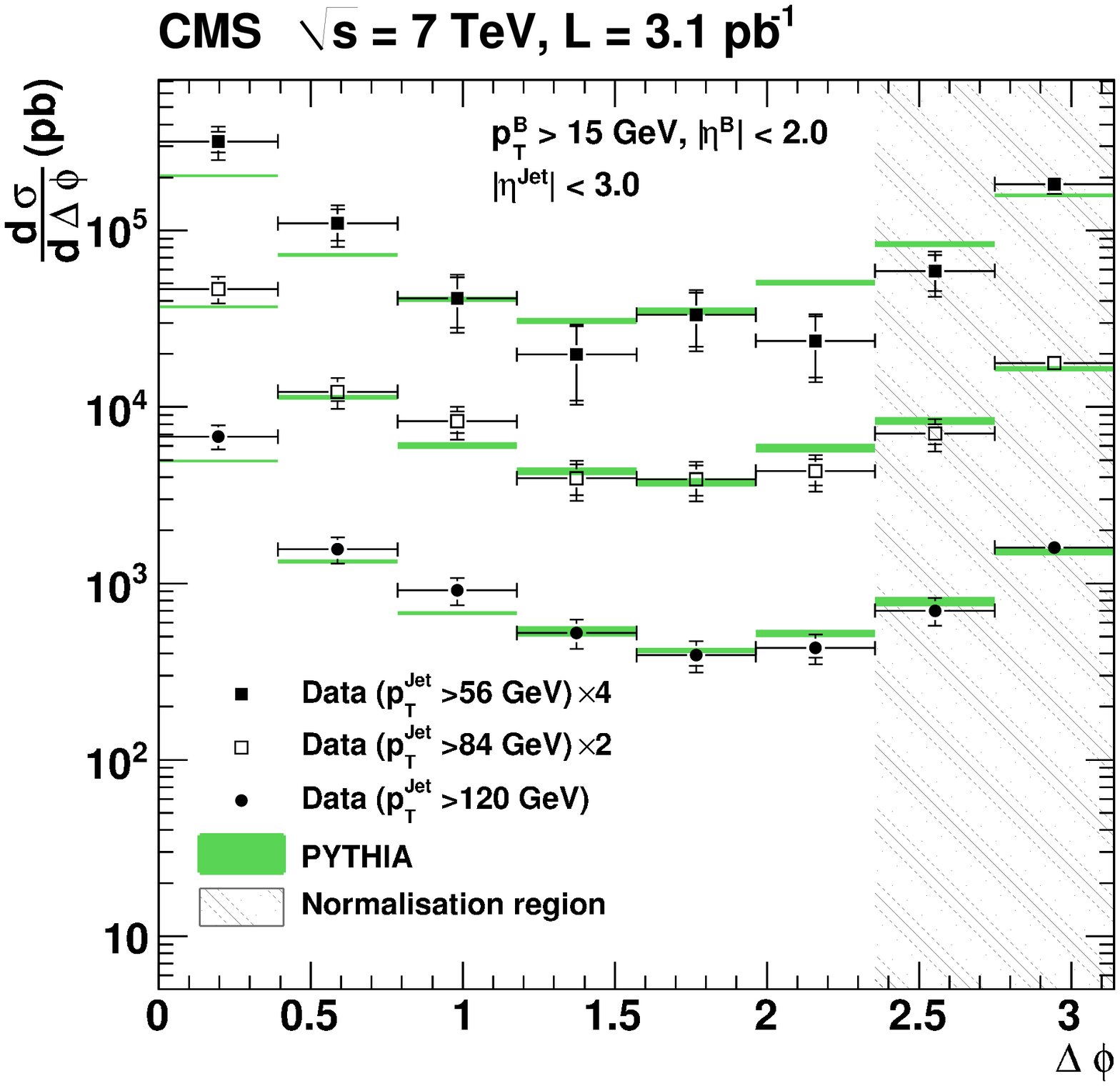}
        \caption{Differential \bhadb production cross sections as a function of 
         $\Delta R$ (left) and $\Delta \phi$ (rad) (right) for the three leading jet \pt\ regions.
   For clarity, the $\pt > 56$ and 84 GeV bins are offset by a factor 4 and 2,              
   respectively. 
   The error bars of the data show the statistical (inner) and the
   total (outer bars) uncertainties. 
   A common uncertainty of 47\% due to the absolute normalisation on 
   the data points is not included.            
   The {\PYTHIA} prediction is normalised to the region $\Delta R > 2.4$ 
   or $\Delta \phi > 2.4$.
   The widths of the shaded bands indicate the statistical uncertainties of the predictions.   
}
\label{fig:b2b}
\end{figure}

The cross sections are determined by applying efficiency corrections
and normalising to the total integrated luminosity.
The angular correlations between the two B hadrons are measured in terms of
the difference in azimuthal angles ($\Delta \phi$) in radians and 
the combined separation variable $\Delta R = \sqrt{(\Delta \eta)^2 + (\Delta \phi)^2}$,
where $\Delta \eta$ is the pseudorapidity.
The analysis results are quoted for the visible kinematic range defined 
by the phase space at the \PB\  hadron level by the requirements
$ |\eta(\mathrm{B})| < 2.0$ and 
$\pt(\mathrm{B}) > 15$ GeV for both of the $\mathrm{B}$ hadrons.
The leading jet used to define the minimum energy scale
is required to be within a pseudorapidity of $|\eta(\mathrm{jet})| < 3.0$.

In order to measure the angular correlations also in the collinear regime,
the reconstruction of the \PB\ hadrons is done independently of jet algorithms.
The method uses the \PB\ hadron decays and is based on an iterative 
inclusive secondary vertex finder
that exploits the excellent CMS tracking information~\cite{Khachatryan:2010pw}.
This allowed to approximate the flight direction of the original \PB\  hadron
by the vector between the primary (PV) and the secondary vertices (SV).
A resolution of 0.02 rad in $\Delta R$ could be achieved that way.
The average overall event reconstruction efficiencies (for both \PB\ hadrons) 
are found to be of order 10\% at an average purity of ~84\%.
Detailed studies were performed to ensure high accuracy in the
B-hadron kinematics description.
In addition, the angular dependence of the efficiency description was verified
by a special event mixing technique, both in data and the simulation.

\begin{figure}[htb]
        \centering
\includegraphics[width=0.45\textwidth]{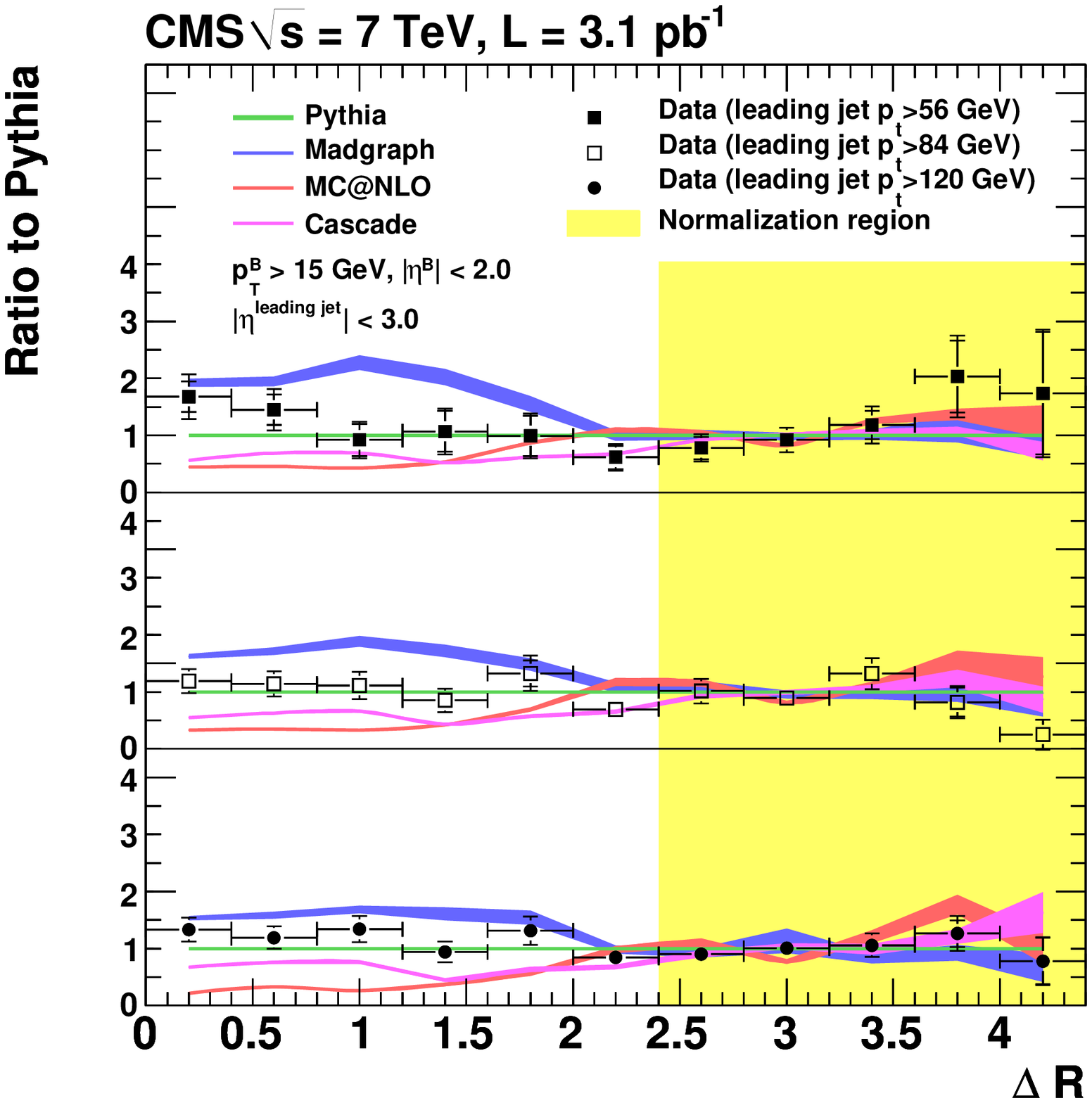}
\includegraphics[width=0.45\textwidth]{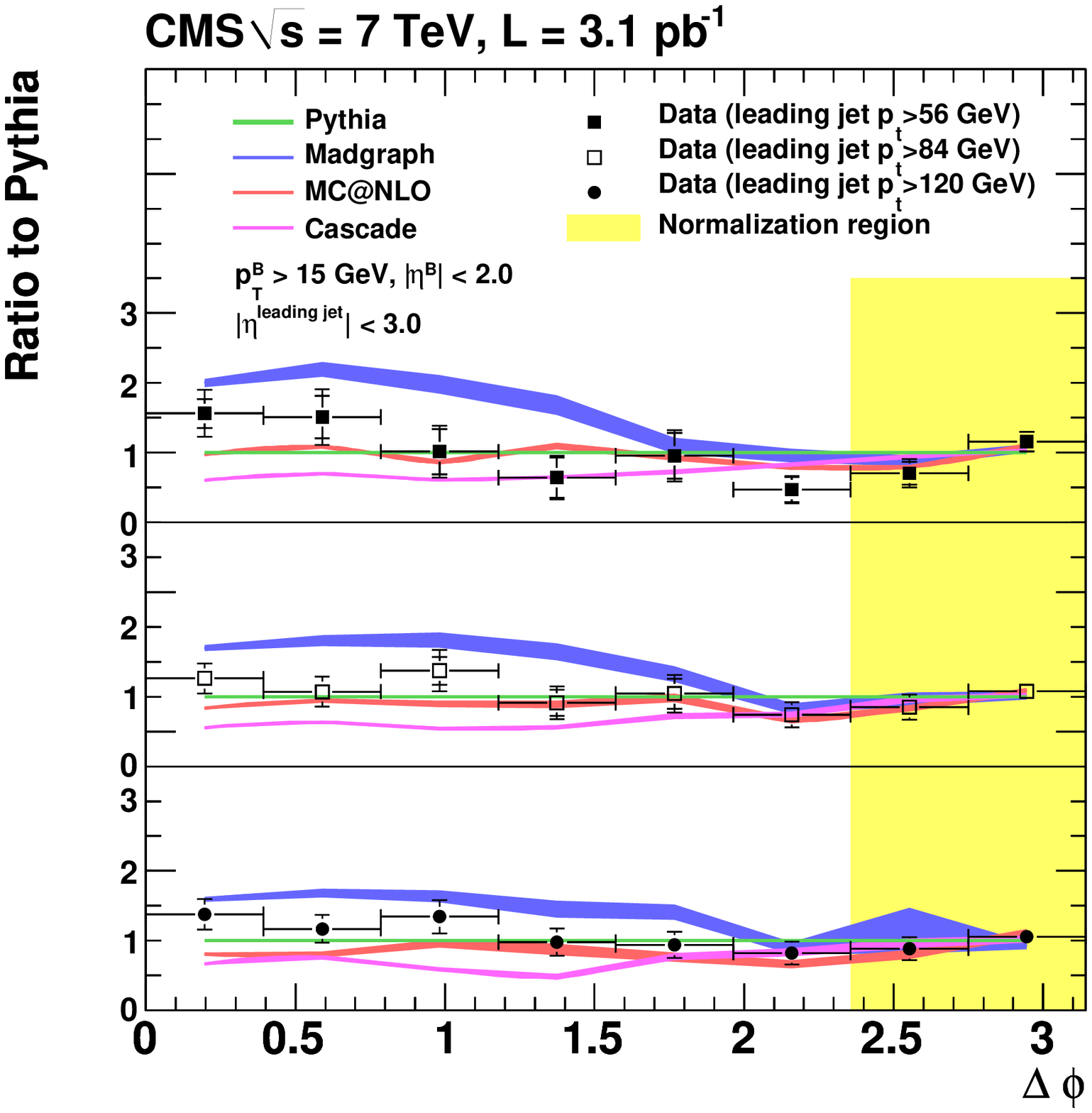}
        \caption{Ratio of the differential \bhadb production cross sections, as a function of       
         $\Delta R$ (left) and $\Delta \phi$ (rad) (right),                                               
         for data, \MADGRAPH, \MCATNLO\ and \CASCADE, with respect to                                
         the {\PYTHIA} predictions, shown also for the three leading jet \pt\ bins.
        The simulation is normalised to the region $\Delta R > 2.4$                                 
        and $\Delta \phi > 2.4$ (rad) (FCR region), as indicated by the shaded normalisation region.
        The widths of the theory bands indicate the statistical uncertainties of the simulation.
}
\label{fig:ratio_b2b}
\end{figure}

The measured cross sections are presented in Fig.\ref{fig:b2b}. 
Overlaid are the predictions
by the \PYTHIA\ calculations, which are normalised to the  $\Delta R > 2.4$  
or $\Delta \phi > 2.4$ (rad) regions, where the calculations are expected to be
more reliable.
Note, that an overall common uncertainty of 47\% 
due to the absolute normalisation is not shown in the figures.

We find that the cross sections at small  $\Delta R$ or $\Delta \phi$ 
are substantial and even exceed the values observed at large angular separation values.
Hence, the configurations where the two \PB\ hadrons are
emitted in opposite directions are much less likely than the
collinear configuration. 

The measurements are compared to various predictions, based on LO and next-to-leading (NLO)
pQCD calculations. Figure~\ref{fig:ratio_b2b} illustrates the shape sensitivity by
showing the ratio of the different $\Delta R$ distributions to the \PYTHIA\
Monte Carlo predictions.
It is found, that the overall tendency in shape is in general reasonably described by the
predictions, however the normalizations and the details in shape, in particular at
small opening angles are not described well by any of the calculations.
Apart from \MADGRAPH\ program, all predictions underestimate the amount of gluon splitting
contributions in the collinear region.

Perturbative QCD predicts a back-to-back configuration 
for the production of the \bhadbbr pair
(i.e.\ large values of  $\Delta R$ and/or  $\Delta \phi$)
for the LO processes. In contrast, 
the region of phase space with small opening angles between
the \PB\  and $\overline{\rm B}$ hadrons provides strong sensitivity to
collinear emission processes, such as the ones present in higher-order
processes. Gluon radiation which splits into
$\mathrm{b\overline{b}}$ pairs 
is anticipated to have a smaller angular
separation between the $\mathrm{b}$ quarks.

The measurements show that the \bhadb production cross section ratio
$\rho_{\Delta R}=\sigma_{\Delta R < 0.8}$ / $\sigma_{\Delta R >2.4}$
 increases as a function of the 
leading jet \pt\ in the event (see Fig.~\ref{fig:gsp_fcr}).
Larger \pt\ values lead to more gluon radiation and, hence, are
expected to produce more gluon splitting into \bhadbbr pairs.
This general trend is described reasonably by the theoretical calculations.

\begin{figure}[htb]
\centering
\includegraphics[angle=90,width=0.45\textwidth]{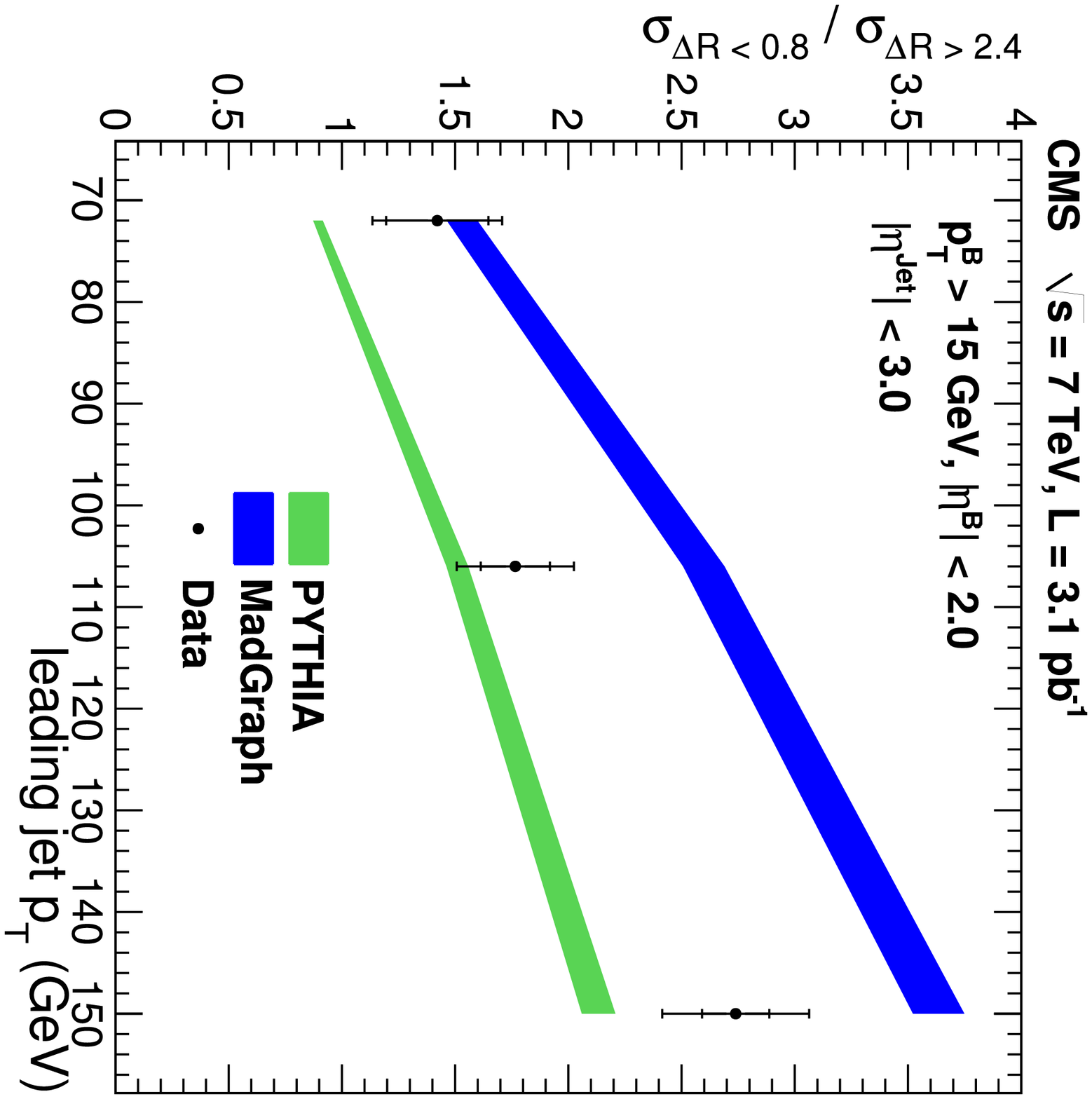}
\includegraphics[angle=90,width=0.45\textwidth]{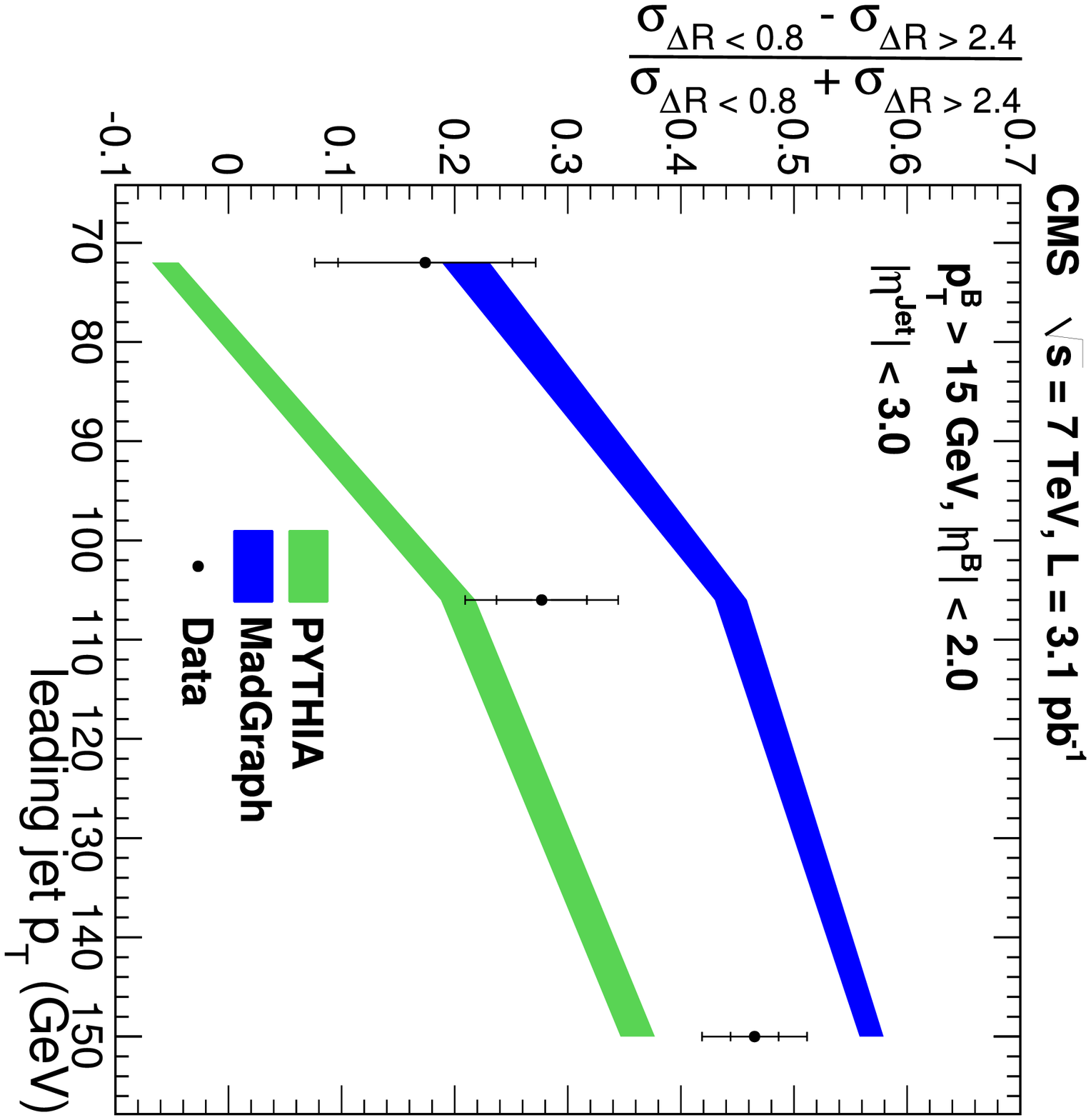}
\caption{Left: ratio between the \bhadb production cross sections in
   $\Delta R < 0.8$ and $\Delta R>2.4$, $\rho_{\Delta R}=\sigma_{\Delta R < 0.8}$ / $\sigma_{\Delta R >2.4}$,
   as a function of the leading jet \pt. Right: asymmetry between the two regions,
$(\sigma_{\Delta R < 0.8} - \sigma_{\Delta R >2.4})$ /
$(\sigma_{\Delta R < 0.8} + \sigma_{\Delta R >2.4})$.
   The symbols denote the data averaged over the bins and
   are plotted at the mean leading jet \pt\ of the bins.
   For the data points, the error bars show the statistical (inner bars) and the
   total (outer bars) errors.
   Also shown are the predictions from the {\PYTHIA} and {\MADGRAPH} simulations,
   where the widths of the bands indicate the uncertainties arising from the 
   limited number of simulated events.
    }
\label{fig:gsp_fcr}
\end{figure}


\section{Conclusions}

The first measurements of inclusive beauty production have been performed at the 
LHC by the CMS experiment over a large range from very low transverse momenta 
up to 300 GeV in the central rapidity region.
Comparisons with theoretical predictions, based on  pQCD calculations
have confirmed the large production cross section.
The calculations in general describe the overall features of beauty production
fairly well. However, the predictions do not yet adequately describe 
the differential measurements, neither in the B transverse momentum, nor the rapidity nor the
\ensuremath{\mathrm{B}}\ensuremath{\mathrm{\bar{B}}} opening angle distributions.



%

}  


\end{document}